\documentclass{aa}
\usepackage{graphicx,natbib}
\bibpunct{(}{)}{;}{a}{}{,}

\def\flux{\rm erg~s$^{-1}$~cm$^{-2}$}
\def\lum{\rm erg~s$^{-1}$}

\begin{document}

\sloppypar

   \title{Large-scale variations of the cosmic X-ray background and
   the X-ray emissivity of the local Universe} 

   \author{M. Revnivtsev \inst{1,2}, S. Molkov \inst{2,1,3}, S. Sazonov \inst{1,2}}

   \offprints{mikej@mpa-garching.mpg.de}

   \institute{
              Max-Planck-Institute f\"ur Astrophysik,
              Karl-Schwarzschild-Str. 1, D-85740 Garching bei M\"unchen,
              Germany,
     \and
              Space Research Institute, Russian Academy of Sciences,
              Profsoyuznaya 84/32, 117997 Moscow, Russia
     \and
              Centre d'Etude Spatiale des Rayonnements, 
              31028, Toulouse, France
            }
  \date{}

        \authorrunning{Revnivtsev et al.}
        \titlerunning{X-ray emissivity of the local Universe}

\abstract{We study the cosmic X-ray background (CXB) intensity variations on 
large angular scales using slew data of the RXTE observatory.
We detect intensity variations up to $\sim2\%$ on angular scales
of 20--40$^\circ$. These variations are partly correlated with the local
large-scale structure, which allowed us to estimate the
emissivity of the local Universe in the energy band 2--10 keV at
$(9\pm4)\times 10^{38}$ \lum\ Mpc$^{-3}$. The spectral energy
distribution of the large-angular-scale variations is hard and is
compatible with that of the CXB, which implies that normal galaxies
and clusters of galaxies, whose spectra are typically much softer, do
not contribute more than 15\% to the total X-ray emissivity of the
local Universe. Most of the observed CXB anisotropy (after exclusion 
of point sources with fluxes $\ga10^{-11}$ erg s$^{-1}$ cm$^{-2}$) can be attributed
to low-luminosity AGNs.
}

   \maketitle

%
%________________________________________________________________

\section{Introduction}
The Universe is a bright source of X-ray emission. The first discovery 
in X-ray astronomy was that of the brightest source in the X-ray sky -- 
the accreting neutron star Sco X-1 -- along with an isotropic cosmic
X-ray background \citep{giacconi62}, which was later shown to
be composed of emission from a large number of discrete extragalactic sources
\citep{giacconi79}. 

Virtually all astrophysical objects produce X-ray emission: starting from
numerous ordinary stars, with typical X-ray (2--10~keV) luminosities
$L_{\rm x}\sim 10^{27}$--$10^{30}$ \lum, through accreting white
dwarfs ($L_{\rm x}\sim10^{30}-10^{34}$ \lum), neutron stars and black holes
($L_{\rm x}\sim10^{35}-10^{38}$ \lum), to active galactic nuclei
(AGNs) and quasars ($L_{\rm x}\sim 10^{40}$--$10^{46}$ \lum).  

The bulk of the cosmic X-ray background (CXB) is created by 
AGNs with luminosities $\sim 10^{42}$--$10^{45}$
\lum\ (e.g. \citealt{ueda03}), but a non-negligible contribution to the
CXB must be provided by low-luminosity ($L_{\rm x}\sim
10^{40}$--$10^{42}$~\lum) AGNs, which constitute the majority of
active supermassive black holes in the local Universe \citep[see
e.g.][]{elvis84,ho97}, as well as by ordinary galaxies, due to their
populations of X-ray binaries and stars. Unfortunately, 
it is difficult to estimate the cumulative luminosities of these
faint extragalactic sources simply by counting them, as is usually
done for powerful AGNs. Indeed, a normal galaxy or a low-luminosity
AGN with an X-ray luminosity $\sim 10^{40}$  
\lum\ located at a redshift typical of CXB sources ($z\sim$1--2) will
produce an X-ray flux less than $10^{-17}$ \flux, which cannot be detected
by existing X-ray telescopes. However, the cumulative emissivity
of all weak sources can in principle be estimated using measurements
of the CXB \citep[e.g.][]{boldt92}. 

It has long been understood that the properties of CXB
sources are not constant but have evolved with cosmic time
\cite[e.g.][]{longair66,schmidt68,silk68,maccacaro83,maccacaro91}. 
In particular, the cumulative X-ray volume emissivity of
AGNs was much higher at redshifts $z\sim$1--2 than it is at the present time
\citep[e.g.][]{maccacaro83,barger05}, the star-formation rate (and
the associated rate of X-ray production) was similarly higher
\citep{madau96}, and so on. It is therefore clear that the integrated 
X-ray emissivity of a unit volume or of a unit mass of Universe has
evolved with time and knowing its local ($z=0$) value is essential for
understanding the evolution of different populations of X-ray sources,
in particular supermassive black holes in the centers of galaxies.

Even the latest all-sky surveys conducted at energies above 2~keV
\citep{revnivtsev04,maretal05,krivonos07}, with their sensitivity
$\sim10^{-11-11.5}$ \flux, can detect sources with luminosities
$\la10^{41}$ \lum\ only at distances smaller than 
10~Mpc, which is not enough to obtain a representative sample of such
objects and to estimate their cumulative luminosity. Stacking of
individually undetectable fluxes from nearby low-luminosity X-ray
sources can hardly help, because the number density of such sources
increases with decreasing luminosity not fast enough ($LdN/d\log
L\propto L^{-(\la 1)}$, see e.g. \citealt{elvis84,persic89}) to compensate for 
the rapidly declining signal-to-noise ratio. 

However, it has been realised that the CXB can help in estimating the
total X-ray emissivity of the local Universe
\cite[e.g.][]{schwartz80,warwick80,
shafer83,jahoda91,miyaji94,carrera95,scharf00}.  
The idea is quite simple -- the Universe has large-scale structure
characterized at zero redshift by a factor of a few or larger density
contrast on scales of tens of Mpc. Since X-ray sources are expected to
trace the overall matter distribution, they should exhibit a similar space
density contrast. When observed in an X-ray survey, the cumulative emission of
all individually undetectable sources located in an overdense region of
the local Universe will create an area of enhanced sky surface
brightness on top of the nearly isotropic X-ray background emission
created at larger distances. The amplitude of this large angular scale
enhancement of the CXB intensity can be converted into a value of the 
cumulative volume emissivity of faint X-ray sources, provided that one
knows the distribution of matter density in that local region of the
Universe and the so-called bias factor of X-ray sources (i.e. the ratio of 
fractional X-ray source number density fluctuations to fractional matter 
density fluctuations). 

Recent advances in the understaning of the bias factor of AGNs indicate
that in the nearby Universe ($z\sim0$) it is close to unity \cite[see
  e.g.][] {tegmark98,boughn04}. Arguably the most direct determination 
of the bias factor (specifically, a measurement of 
the ratio of fractional number density variations of X-ray sources to
fractional number density variations of IRAS infrared-selected
galaxies) was recently made using the INTEGRAL all-sky hard X-ray
survey (Krivonos et al. 2007, in preparation), and the AGN bias factor was
confirmed to be close to unity. This implies that to convert CXB
intensity variations into an X-ray emissivity of the local Universe we can
use any good tracer of mass density 
in the local volume, for instance infrared galaxies \cite[see
  e.g.][]{basilakos06}. Supporting evidence that AGNs closely trace the
distribution of galaxies was recently presented by
\cite{li06} based on an optically selected sample of narrow-line AGNs from
the Sloan Digital Sky Survey.

According to the current knowledge of the distribution of matter in the local
Universe, even its largest angular scale anisotropy -- the dipole
component -- continues to grow to distances $\ga100$--200~Mpc 
\cite[e.g.][]{rowan90,kocevski04,basilakos06}, where 
the distribution of matter is not yet well known. Nonetheless, there are
distinct density patterns in the nearest ($\la 30$--70~Mpc) Universe
that should be imprinted into the CXB intensity distribution and
could be studied with the available X-ray data. 

In this paper we essentially repeat the analysis of large-scale
anisotropies of the CXB, which was previousely performed based on HEAO1/A2 data
\citep[e.g.][]{shafer83,jahoda91,miyaji94,scharf00,boughn02},  
using a newly available dataset from the PCA instrument aboard RXTE and
redetermine the X-ray emissivity of the local ($z=0$) Universe utilizing the
advantages of the higher collective area of RXTE/PCA and its smaller
field of view (and consequently the reduced confusion noise). 

Thoughout the paper we adopt a Hubble constant $H_0=75$ km 
s$^{-1}$ Mpc$^{-1}$ and recalculate all relevant quantities from cited 
papers to this reference value.

\section{Data analysis}
\label{sec:analysis}
\subsection{Systematic limitations}
\label{subsec:systematics}

The high level of isotropy of the cosmic X-ray background was noticed already 
shortly after its discovery \cite[e.g.][]{schwartz70}. In order to 
detect and accurately measure the anisotropy of the CXB caused
either by our motion with respect to the outer Universe where the
bulk of the CXB is created (the so-called Compton--Getting, or CG
effect, \citealt{compton35}) or by the large-scale structure of the Universe
at zero redshift one should reach an accuracy of measuring the CXB intensity
of 0.1--0.5\% of its average level.

Indeed, the relative dipole-like ($\delta I/I\propto 1+\Delta \cos
\theta$) variations of the CXB intensity due to our motion with respect to 
the outer Universe radiation field (the cosmic microwave background, CMB)
are expected to be \cite[see e.g.][]{scharf00} on the order of

$$\Delta_{\rm CG}=(v/c)(\Gamma+2)\sim 4.2\times10^{-3},$$
where $\Gamma\approx1.4$ is the photon index of the CXB in our RXTE
working energy band 3--20 keV. 

The X-ray dipole due to large-scale structure in the local Universe,
produced by inhomogeneities in the matter density distribution at
distances $D_{\rm LSS}\la$50--100 Mpc, is expected to be

$$
\Delta_{\rm LSS}\sim {\phi_{\rm z=0} D_{\rm LSS} \over{4\pi I_{\rm
CXB}}} \sim (0.6-1) \times 10^{-2}, 
$$
where $\phi_{\rm z=0} \sim 10^{39}$ \lum\ Mpc$^{-3}$ is the average
volume emissivity of the local Universe and $I_{\rm CXB}\sim 2\times
10^{-11}$~erg~s$^{-1}$~cm$^{-2}$ is the average CXB intensity in the 2--10~keV
energy band.

Such accuracy presents a challenge for a vast majority of X-ray 
instruments. One of the best instruments suitable for measuring the
tiny large-scale CXB anisotropies was the A2 experiment aboard the HEAO1
observatory \citep{heao1a2}. It had a special design of detectors that
made it possible to subtract the internal background of the detectors
with almost absolute accuracy. However, its detectors had relatively
large fields of view ($1.5\times3$ deg, $3\times3$ deg and $3\times6$
deg) and therefore seriousely suffered from confusion noise produced
by sources with flux below the detection threshold.

The field of view of the RXTE/PCA spectrometer is significantly
smaller ($\sim 0.974$ sq.~deg), while its effective area 
is significantly larger ($\sim 6400$ cm$^2$ of all the detectors combined, 
or $\sim1300$ cm$^2$ of a single detector), and the model of the PCA
detector background developed by the PCA team allows one to subtract it with
an accuracy of at least 2--3\% of the CXB intensity \citep{jahoda06}. 
Therefore, RXTE/PCA is currently the best instrument for
studying large-scale anisotropies of the CXB.

The noise created by point sources with flux just below the
detection threshold  
is still an issue for RXTE/PCA. Its influence can be estimated
as follows \cite[see e.g.][]{warwick80}. Suppose that the differential
number-flux function of point sources $dN(S)$ is isotropic over the sky and 
can be described by a power law with a slope of $-5/2$:

$$
dN(S)=\Omega \, A\, S^{-5/2} dS,
$$
where $S$ is the source flux, $A$ is a normalization constant and
$\Omega$ is the solid angle subtended by the survey.

The number of sources in any infinitely small flux bin is subject to
Poisson statistics, hence we can write for the dispersion of the integrated
signal: 

$$
(\delta I)^2 = \int_{S_{\rm min}}^{S_{\rm max}}{S^2
  {dN\over{dS}} dS} \approx 2\Omega A S_{\rm max}^{1/2},
$$
where $S_{\rm max}$ is the maximal flux of undetected sources.

The relative uncertainty in the flux from the solid angle $\Omega$
will be $\delta I/I \propto S^{1/4}_{\rm max}/\Omega^{1/2}$. For the
real number-flux function of extragalactic sources in the energy band
2--10 keV \cite[$A\approx10^{-2}$ deg$^{-2}$ $(10^{-11}$ erg/sec/cm$^2$)$^{-1}$, see e.g.][]{revnivtsev04} this will be
\begin{equation}
\left({\delta I\over{I}} \right)_{\Omega}\sim 8.7\times 10^{-2} {S_{\rm max,11}^{1/4} \Omega_{\rm deg} ^{-1/2}}.
\label{poiss}
\end{equation}
Here $S_{\rm max,11}$ is the maximal flux of sources 
in units of $10^{-11}$ \flux, and
$\Omega_{\rm deg}$ is the solid angle of the survey in degrees. We
adopted the CXB intensity to be equal to $1.95\times10^{-11}$ 
\flux\ deg$^{-1}$ \citep[][see also more on this
  below]{revnivtsev03,revnivtsev05}.  

From the above estimates it is clear that in order to reach the
required accuracy of (1--5)$\times 10^{-3}$ of the CXB intensity one
needs to exclude sources down to a flux limit of $\sim 10^{-11}$ \flux\ and 
average CXB intensity measurements over an area of 400--1000 deg$^2$.
In reality the number-flux function of extragalactic sources flattens
below $10^{-14}$ \flux\ \citep[e.g.][]{hasinger93,vikhlinin95}, but this
does not strongly change the above estimates.  

\subsection{RXTE/PCA slew data}

Over its lifetime (Dec 1995--present), the spectrometer PCA on the
RXTE observatory has collected a lot of data during slews between
pointed observations. The high collective area of the PCA detectors
($\sim 1300$ cm$^2$ of each of the five) allows one to detect sources down to
fluxes $\sim5\times10^{12}$ \flux\ and this detection limit is
determined solely by the confusion noise \citep{markwardt04,revnivtsev04}.

For our analysis we used all publicly available data of RXTE/PCA scans and 
slews from February, 1996 till February~8, 2007. During a significant period
of RXTE operations only some of the five PCA detectors were
switched on (this mode of operations became dominant after the end of Epoch3 of
the PCA voltage, i.e. since March 1999), therefore not all of the PCA
detectors have collected sufficient exposure time and sky coverage for
our purposes. For this reason, in our analysis we used only detector
\#2, which collected $\sim$6--6.5~Msec of data between February, 1996
and February~8, 2007 (we consider here only regions away from the 
Galactic plane, $|b|>14^\circ$, and $>3^\circ$ away from the detected
point sources reported in \citealt{revnivtsev04}). 

We analized the data of the first layer (which typically enables the highest 
signal-to-noise ratio for weak X-ray sources) of the PCU2 in two
energy bands, 3--8 and 8--20 keV, which were converted into
instrumental channels using response matrices (constructed with the HEASOFT 6.0
package) appropriate for each observation. We similarly analyzed
observations of the Crab nebula, 6 background pointings  
(see \citealt{jahoda06}) and ``dark'' Earth\footnote{An observation was
considered to be pointed at dark Earth if the angle between the
vector Sun--Earth and the vector directed from the center of Earth
towards the point on its surface observed by RXTE/PCA is larger than
$90^\circ$ and the RXTE/PCA field of view is directed more than
$10^\circ$ below the Earth's limb.}.  

Flux measurements by RXTE/PCA are subject to systematic uncertainties. 
Absolute flux measurements (in particular those of the CXB) 
depend on how well the effective area of the detector
is known, the knowledge of the effective solid angle of the collimator
and so on. In our analysis we relied on an absolute X-ray flux
calibrator --  the Crab nebula. Specifically, our results are obtained
under the assumption that the Crab spectrum is $dN/dE=10
E^{-2.05}$ (which implies a flux in the 2--10 keV energy band $F_{\rm
  Crab}=2.39\times10^{-8}$ \flux). 

As we are interested in small X-ray intensity variations over the sky
we should take into account all systematic uncertainties that could 
affect our measurements at the level of a per cent or so.

Our data analysis consists of the following steps:

\begin{itemize}

\item Background subtraction using standard HEASOFT 6.0
tasks. Aggresive filtering of PCA data: based on electron counts
({\tt ELECTRON$<$0.08}), Earth elevation angle ({\tt ELV$>$10}), of
3$^\circ$-radius regions around known detected point sources   
\citep{revnivtsev04} and new source candidates (we have filtered out
all sky cells of $\sim1.5^\circ\times 1.5^\circ$ size where more than 
5$\sigma$ count rate excesses were detected), and of a wide band around
the Galactic plane (see more on this in Sect \ref{galaxy}).

\item Correction for the small ``unmodelled'' part of the RXTE/PCA
background.

\item Correction for PCA detector parameters' drifts based on measurements
of the Crab count rates in the considered energy bands.

\item Additional check of the accuracy of all corrections by studying the
time dependence of the flux measured from dark Earth, which in the 
case of the RXTE/PCA background model $=-$ the CXB flux. Here
we assume that dark Earth emits zero X-ray flux in the studied
energy band, because the albedo of the Earth atmosphere is small at these
energies (even though it may be significant at energies above
15--20~keV) \citep{churazov06} and the cosmic ray-induced
atmospheric hard X-ray radiation is still highly photoabsorbed at
these energies \citep{sazonov07}. In addition, we check the time dependence  
of the contrast between fluxes collected from spatially distinct
regions of sky.

\item Estimation of the contribution of the Galaxy to the sky X-ray
intensity at high Galactic latitudes. 

\end{itemize}

Below we describe these steps in more detail.

\subsection{Background subtraction}

One of the main systematic uncertainties of relative flux measurement
is the accuracy of the PCA background subtraction. 
The currently standard approach to this consists of modeling the
background with the highest possible accuracy using 
a number of tracers recorded onboard on the 16-sec time scale and
verifying that ``empty'' sky regions have zero intensity after background 
subtraction. Although the CXB contribution to the flux measured by RXTE/PCA is
eliminated by this procedure, it can be recovered by observations of
dark Earth \cite[see e.g.][]{revnivtsev03}.  

\begin{figure}
\includegraphics[width=\columnwidth]{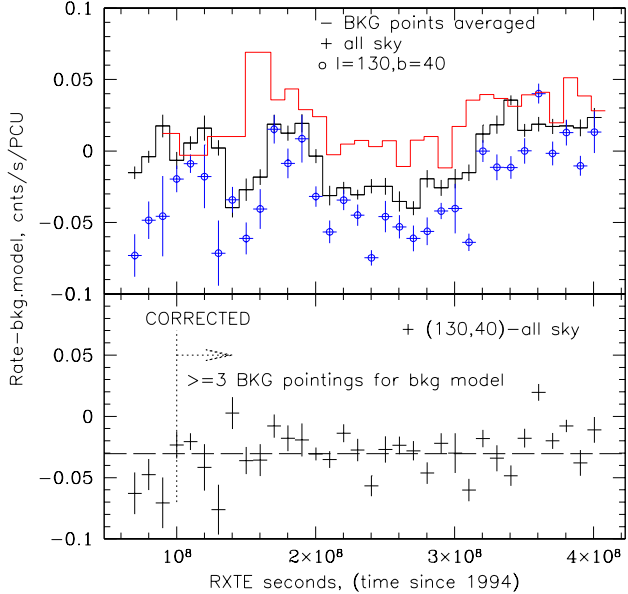}
\caption{{\it Upper panel}: background subracted intensities of sky, averaged
over 4-month intervals. Crosses denote all-sky averaged values, open
circles the intensity integrated over a $25^\circ$-radius circle around
$(l,b)=(130^\circ,40^\circ)$ and the solid line the intensity averaged
over the 6 RXTE/PCA background pointings. It can be seen that all three curves
demonstrate similar variations with time, which we consider the
unmodelled part of the RXTE/PCA instrumental background. {\it Lower
panel:} background subtracted intensity from
$(l,b)=(130^\circ,40^\circ)$ with the  
all-sky-averaged value subtracted. The unmodelled intensity variations
have become much smaller and are almost compatible with the
statistical uncertainties.}
\label{bkg_patterns}
\end{figure}

The current version of RXTE/PCA software (LHEASOFT 6.0) provides
an outstandingly accurate model of the instrumental  
background ({\tt L\_7\_240CM} version of the background), which allows for an 
accuracy of background subtraction of a few per cent, i.e. the
root-mean-square amplitude of the unmodelled deviations is
$\sim$0.02--0.03 cnts/s/PCU/beam in the 3--8 keV energy band, which
corresponds to $\sim 3\times 10^{-13}$ \flux\ deg$^{-2}$ for a
Crab-like spectrum \citep[e.g.][see also
  Fig.~\ref{bkg_patterns}]{craig02,jahoda06}, or $\sim2$--3\% of the 
average CXB intensity. 

However, as our aim is to measure the CXB intensity with an accuracy
of 0.1--0.5\%, we made special corrections to take into account
the unmodelled part of the RXTE/PCA background, taking advantage of
the observational fact that this part appears to be an additive
component to all measured count rates irrespective of the pointing of
a given PCA observation. As an example of such a behavior, we show in
Fig.~\ref{bkg_patterns} background subtracted, 4-month-binned count
rates averaged over the whole sky, over the cone of $25^\circ$ half-opening
angle around $(l,b)=(130^\circ,40^\circ)$ and over the 6 RXTE/PCA
pointings that were used to calibrate the PCA background model. The
similarity of the patterns of the unmodelled parts of the background
in these three cases is apparent. 

If we assume that the true intensity averaged over the whole sky is zero
at relatively high Galactic latitudes $|b|>25^\circ$ (to eliminate the 
contribution from the Galactic ridge X-ray emission, see Sect.~\ref{galaxy})
 and also excluding $3^\circ$ regions around detected point sources,
the 4-month-averaged measurements of sky intensity provide us an
estimate of the unmodelled part of the RXTE/PCA background, which we
add to the model-predicted rates of the PCA background.

In our subsequent analysis we only used data obtained later than 
$10^8$ {\tt RXTE seconds} (i.e. after March 3rd, 1997), 
because only during this period more than 
3 background poinings were used to calibrate the PCA background model, 
which apparently has an effect on the accuracy of the background subtraction 
(see Fig.~\ref{bkg_patterns}).

The effect of the additional background subtraction is demonstrated in
the lower panel of Fig.~\ref{bkg_patterns}. We see that the
two-step background subtracted sky intensity from some sky region 
is much more stable than that obtained after just the first (standard)
step and that the variations of the inferred intensity are almost
compatible with the statistical uncertainties of the measurements. 

\subsection{Drift in the detector parameters}

Another important issue in X-ray flux measurement is the parameters
of the detector. The efficiency of the PCA detectors slightly changed
during the large time span considered here (1996--2007) for different
reasons, including slow drift of the energy scale of the 
detectors and a slight change of the gas purity or pressure withing
the signal volumes \cite[see e.g.][]{jahoda06}.

\begin{figure}
\includegraphics[width=\columnwidth]{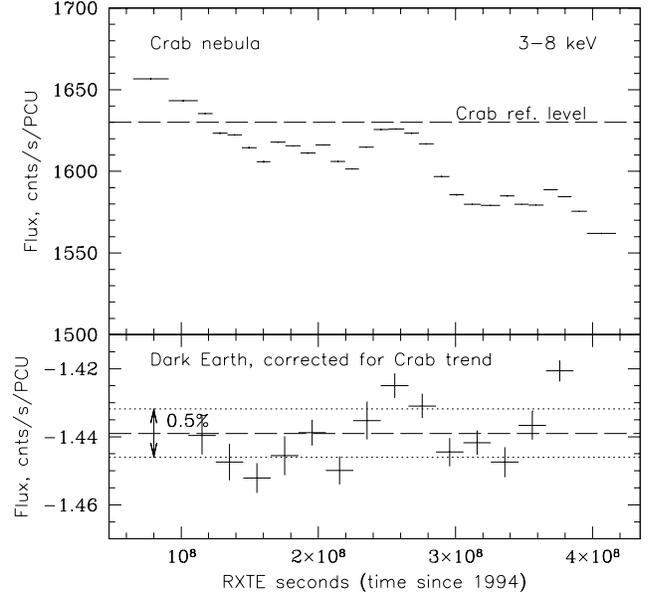}
\caption{{\em Upper panel:} background subtracted count rate 
of the Crab nebula measured in the energy band 3--8 keV with RXTE/PCA
over its lifetime. The count rate is not corrected for the deatime
fraction which is typically $\sim4$\% higher during Crab observations
than during off-source observations. {\em Lower panel:}
Background subtracted count rate of dark Earth corrected for the
``unmodelled'' background and for the drift of the detector parameters.
}
\label{crab_earth}
\end{figure}

This small ($\sim5$\%) drift of the detector effective area manifests
itself in the measured count rates from the Crab, a constant X-ray source
which has been regularly observed by RXTE (see Fig.~\ref{crab_earth}). We
made a special correction for this effect, modeling it by a piecewise
function.

In order to check the stability of the corrected count rates we
studied the detector count rate during dark-Earth observations. By
construction, the PCA background model eliminates the count rate
during off-bright-source observations, automatically subtracting  
the average CXB flux contribution. Observations of dark Earth 
therefore allow one to see this oversubtracted part of the PCA
background, which is essentially the all-sky average CXB flux
\cite[see][]{revnivtsev03}. The count rates measured during dark-Earth
observations, corrected for all the above effects, are shown 
in Fig.~\ref{crab_earth}. We see that the resulting dark-Earth count
rate ($=-$CXB flux) is stable to within $\sim$0.5--1\%. 

Similar corrections were done in the energy channel 8--20 keV.

\begin{figure}
\includegraphics[width=\columnwidth,bb=10 386 570 700,clip]{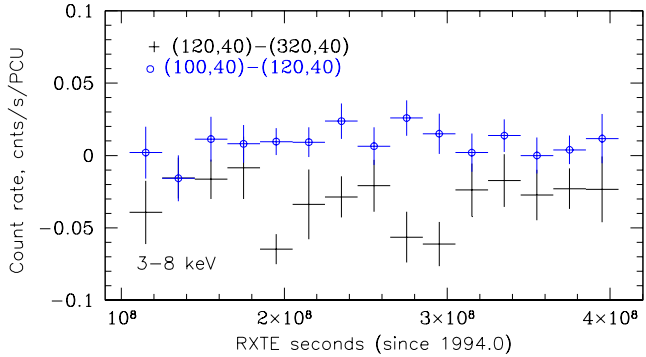}
\caption{Differences of the count rates integrated over a 
$25^\circ$-radius cicle around different $(l,b)$ positions at various
epochs (approximately 4-month averaged): between
$(120^\circ,40^\circ)$ and $(320^\circ,40^\circ)$  -- crosses, and 
between $(100^\circ,40^\circ)$ and $(120^\circ,40^\circ)$  -- open
circles. It can be seen that these differences are quite stable with
time and apparently not strongly affected by systematic effects. The
uncertainty caused by the Poisson 
variations of the number of unresolved point sources within the considered
regions (see eq.~\ref{poiss}) is $\sim 1.5\times10^{-3}$ cnts/s/PCU.}
\label{compare_two_regions}
\end{figure}

After all the corrections, the CXB intensity measured within relatively
large sky areas (so that the Poisson variance of the number of unresolved
sources within a given area does not influence the result) exhibits 
definite variations over the sky. We studied the time dependence of
these variations to check for any remaining systematic effects in our
data. The difference in the CXB intensity between various sky regions
revealed no systematic variations ($\sigma_{\rm 
 systematic}<0.01$ cnts/s/PCU, see Fig.~\ref{compare_two_regions}).  

\subsection{CXB intensity}
As a byproduct of the analysis described above we have estimated the
CXB intensity in two broad energy channels ($=- {\rm dark~Earth~flux}$). The
averaged CXB fluxes measured by PCA in the bands 3--8 keV and 8--20 keV
are $0.847\pm0.008$ mCrab/beam and $1.43\pm0.05$ mCrab/beam,
respectively. Note that in making these estimates we adopted that the
average deadtime fraction during Crab observations is $\sim4$\% higher
than that during empty sky/dark Earth observations \cite[see e.g.][]{jahoda06}.

For a power-law spectral shape of the CXB with a photon index
$\Gamma=1.4$ the flux measured by PCA in the energy band 3--8 keV
corresponds to a flux at 2--10 keV of
$(1.95\pm0.02)\times10^{-11}$ \flux\ beam$^{-1}$. Taking into account
that the effective solid angle of the PCA spectrometer is $\sim0.974$
sq.~deg, we find that the average CXB intensity $I_{\rm
  CXB,~2-10~keV}=(2.00\pm0.02)\times10^{-11}$ \flux\ deg$^{-2}$.

\subsection{Contribution of the Galaxy}
\label{galaxy}

Our Galaxy is a bright emitter in soft \citep[e.g.][]{mccammon90}, standard 
\citep[e.g.][]{worrall82,revnivtsev06} and hard X-rays
\cite[e.g.][]{krivonos07}. Due to the large angular scale of this
emission, it can contribute to the large-scale anisotropy of the CXB
and therefore this contribution should be estimated and subtracted or masked.

The operating energy band of RXTE/PCA (3--20 keV) precludes a
detection of the soft X-ray background of the Galaxy, hence only
the Galactic ridge X-ray emission (GRXE) is important for our study. 

It was recently shown that most likely the GRXE is a
superposition of weak Galactic sources, namely cataclysmic variables
and active stars, and that its surface brightness closely follows the
near-infrared (NIR) surface brightness of the Galaxy \citep{revnivtsev06}. 
Therefore, one can estimate the ridge X-ray intensity in a
given direction by rescaling the NIR map of the Galaxy.

However, due to the discrete nature of the GRXE one should not expect to see 
this emission in regions of extremely small surface number density of stellar
objects, such as the Galactic poles. Indeed, if we assume the volume
emissivity distribution of Galactic X-ray emission to be $\phi(z)=dL_{\rm
  x}/dV\sim 1.5\times 10^{26} \exp(-z/z_0)$ \lum pc$^{-3}$ with the exponential
scale hight $z_0\sim130$~pc \citep{revnivtsev06}, the
flux from a solid angle $\Omega$ in the direction of the
Galactic poles ($|b|=90^\circ$) will be
$$
F=\int^{\infty}_{0}{ {{\phi(z)\Omega  z^2 dz}\over{4\pi z^2}}}\sim2\times10^{-9} {\Omega\over{4\pi}} {\rm erg \,\, s^{-1}\, cm^{-2}}.
$$
For a 1~sq.~deg solid angle the flux will be $\sim5\times10^{-14}$ \flux, i.e
$\sim0.25$\% of the CXB flux. Some addition from stellar 
population of spheroid of our Galaxy \citep{bahcall86} can be anticipated, 
however it should not be larger than that of the disk.

\begin{figure}
\includegraphics[width=\columnwidth]{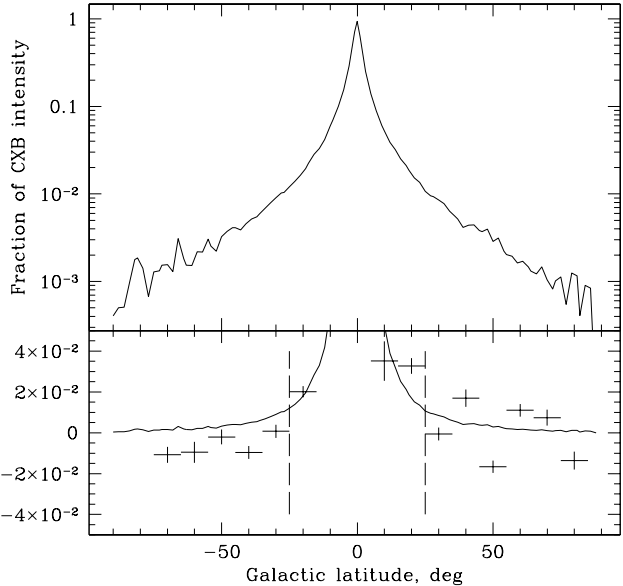}
\caption{{\sl Upper panel:} Latitudinal profile of X-ray (3--8~keV) surface 
brightness of the Galaxy, obtained by rescaling the NIR map of the sky 
\citep{revnivtsev06}, presented in units of the average CXB intensity
in the 3--8~keV energy band. The value of the NIR sky brightness at the
Galactic poles has been subtracted. The small variations of the NIR
intensity at $|b|>50^\circ$ are due to weak point sources present in
the studied areas. It can be seen that at $|b| > 25^\circ$, the
contamination of the CXB by the Galaxy is smaller than $\sim1$\%. In
reality it should drop much faster at $|b|>20^\circ$ due to the
discrete nature of the GRXE (see text). {\sl Lower panel:} The
same profile is shown in linear scale along with real measurements of
the X-ray (3--8 keV) intensity at $|l|<40^\circ$. All the flux
measurements were corrected following the prescriptions described 
in Sect.~2. The dashed lines denote the area ($|b|<25^\circ$) that was
excluded from the subsequent analysis of the CXB maps due to the apparent
non-negligible contrbution of Galactic X-ray emission.}
\label{profile_l40}
\end{figure}

However, 
in reality the GRXE flux from these directions should be even 
lower because sources creating the GRXE are expected to be quite rare
there. Indeed, in the direction of the Galactic poles the mass of
stars in the solid angle $\Omega$ is (here we assume for simplicity an
exponential distribution of stellar density in the Galactic disk,
$\rho(z)=\rho_0 \exp(-z/z_0)$, the stellar density in the Solar vicinity being
$\rho_0=0.04 M_\odot$/pc$^3$, \citealt{robin03})

\begin{eqnarray*}
M(b) &=& \int_{0}^{\infty}{{\rho_0 e^{-z/z_0} \Omega z^2\over{|\sin^3 b|}}
  dz}={2 \rho_0\Omega z_0^3\over{|\sin^3 b|}} \\
&\sim& 50 M_\odot~{\rm for}~|b|=90^\circ~{\rm and}~\Omega=1~{\rm sq.~deg}.
\end{eqnarray*}
Such a small stellar population will simply not contain any X-ray
sources, such as cataclysmic variables, which create the bulk of the
GRXE at energies 3--20 keV. Indeed, more than 50\% of the GRXE is
created by sources with luminosities $>10^{30}$ \lum\
\citep{sazonov06,revnivtsev07}, whose density is  
smaller than $10^{-3} \, M_\odot^{-1}$. Therefore, more than $M_{\rm lim}
>10^{3} M_\odot$ of stars are needed to create significant X-ray flux
from the studied direction. Such a mass in a $\Omega\approx1$~sq.~deg solid
angle will be present if $|\sin^3(b)|\la 2\Omega \rho_0 z_0^3/M_{\rm
  lim}$, i.e. for $|b|\la 20^\circ$. Nonetheless, for rough estimates of the
GRXE contribution one can use the scaled (as shown in
\citealt{revnivtsev06}) brightness of the Galaxy at 3.5~$\mu$m (see
Fig.~\ref{profile_l40}). 

After all filterings we constructed an X-ray intensity profile of the sky 
along the Galactic latitude, by averaging measured intensities in
$|l|<40^\circ$-wide longitudinal bins (Fig.~\ref{profile_l40}). 
At latitudes $|b|>25^\circ$ no dependence of the sky surface
brightness on latitude exceeding the involved uncertainties is seen,
hence we mask the wide $|b|<25^\circ$ region from our subsequent analysis.

We note that the signal previousely seen at $25^\circ<|b|<60^\circ$ by 
HEAO1/A2 \citep{iwan82} may be created by unresolved Galactic and
extragalactic sources for which the HEAO1/A2 instrument was less
sensitive than PCA due to the wider field of view of the former and
consequently higher confusion noise. This is supported by the 
fact that \cite{iwan82} claimed the Galactic background emission to be
present at $|l|>90^\circ$ at a level comparable to that at
$|l|<90^\circ$, while the latest studies show that there is virtually
no GRXE emission from the Galactic anticenter region \citep{revnivtsev06}.

\section{Results}

\subsection{Large-scale structure of the local Universe}

The Universe is highly homogeneous on large scales ($\gg
100$--200~Mpc, e.g. \citealt{wu99}). Due to this uniformity and
because most  of the CXB originates at relatively large distances from 
us (at redshifts $z\sim1$--2, e.g. \citealt{ueda03,hasinger05}), the CXB is very
isotropic. However, at smaller distances the Universe becomes
inhomogeneous and this should cause some anisotropies of the CXB
intensity. Since the CXB is composed of discrete sources, these
anisotropies should be studied on angular scales sufficiently 
large that the Poisson variations of the number of sources within the
area of study are unimportant, as described by equation~(\ref{poiss}). 

The dipole component of the CXB intensity caused by the local large-scale
structure saturates at distances $D\la200$--300~Mpc
\citep[e.g.][]{scharf00, kocevski04}. However, higher order harmonics
might be created by more distant structures. This fact can
significantly complicate the comparison of the CXB anisotropies with
the structures in the nearby Universe known from sky surveys 
\citep[e.g. the well-known IRAS PSCz survey in the infrared
band,][]{saunders00}, because typically such structures are
well probed only out to distances $\sim$100--200~Mpc. 

In order to study the amplitude of CXB anisotropies that may be created
by more distant structures we analyzed the distribution of 
galaxies resulting in the semi-analytic modeling of formation of
galaxies in Millennium simulation
\citep{delucia06}. It was shown \cite[e.g.][]{wang07} that the distribution of
$z\approx0$ galaxies in this simulation has statistical properties
resembling those of the real populations of galaxies in the local
Universe. Our examination of the distribution of simulated galaxies
showed that the matter density structures at distances $>100$--200~Mpc create 
fluctuations of the CXB intensity comparable in amplitude with those
created by the more nearby mass concentrations. Therefore, one should
not anticipate the CXB intensity map to resemble a map of the
nearest mass concentrations only. 

\subsection{CXB intensity maps and dipole component}

\begin{figure}
\includegraphics[width=\columnwidth]{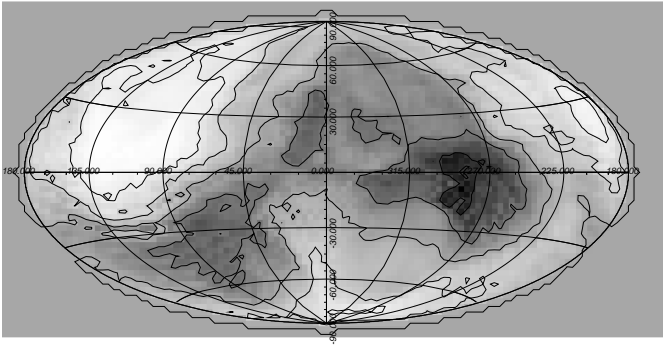}
\caption{Adaptively smoothed map of the CXB intensity in the 3--8 keV
energy band, in Galactic coordinates. Plotted is the X-ray intensity averaged
over cones with open angles 20--40$^\circ$, excluding areas around point
sources and a band around the Galactic plane ($|b|<25^\circ$). The
contours represent the levels of deviations of the CXB intensity from
its average value by -1\%, -0.5\%, 0\%, 0.5\% and 1\%.} 
\label{map3_8}
\end{figure}

After implementing all the masks and corrections (see the previous
Section), we constructed maps of sky intensity (with the average
CXB intensity subtracted as a consequence of the PCA background
subtraction) in the energy bands 3--8 keV (Fig.~\ref{map3_8}) and 8--20 keV. 

The simplest all-sky anisotropy that can be studied using these maps
is the CXB dipole component. 

A simple determination of the dipole component of the CXB angular 
distribution -- its approximation by the function  $I/I_0=1+\Delta
\cos\theta$ (where $\theta$ is the angle between the given direction and
the direction of the dipole) -- yields the amplitude $\Delta$. The obtained
parameters of the CXB dipole in the 3--8~keV energy band are presented
in Table \ref{dipoles}. 

Several effects can produce a dipole component of the
cosmic X-ray background. First, it may arise from some primordial anisotropy 
of the Universe at high redshifts.

 Second, the observed CXB
dipole must contain a contribution from the Doppler effect associated
with our motion with respect to the reference frame in which the bulk
of the CXB emission is created ($z\sim1$--2). 

Finally, relatively
nearby ($z<0.1$--0.5) mass concentrations can impose their imprints on
the CXB map, and in particular create a non-negligible dipole
component. According to our current understanding of the distribution
of matter density in the nearby Universe, it indeed contains a dipole
component that causes a bulk motion of the local group of galaxies,
which in turn can be seen via measurements of the dipole component in
the map of the cosmic microwave background \cite[see e.g.][]{rowan90}.

As dipole anisotropy is a very wide feature on the sky, it is very hard
to disentangle in the CXB intensity map the different components of the total
dipole signal described above. 

One promising approach would be to study the
dependence of the amplitude of CXB intensity variations on
energy. Indeed, as was already mentioned in
Sect.~\ref{subsec:systematics}, the amplitude of the Compton--Getting 
dipole component (caused by our motion with respect to
the CXB reference frame) has a strong dependence on the slope of the
CXB spectrum in the considered energy band: $\delta I/I\propto 
\Gamma(E)+2$, where $\Gamma(E)$ is the characteristic power-law photon
index of the spectrum at energy $E$. This means that in hard X-rays
(50--200 keV), where the CXB spectrum has a much steeper slope
$\Gamma\sim2.2$--2.6 than at energies below $\sim 20$ keV
($\Gamma\sim1.4$), the amplitude of the CG effect must be considerably 
higher and have a specific energy dependence. On the other hand,
the spectrum of the CXB dipole component created by large-scale structures 
in the nearby Universe is expected to be similar to that of the cumulative
emission spectrum of local X-ray sources (mainly AGNs), 
which resembles the CXB spectrum blueshifted by a factor of $\sim 2.5$
\citep{sazonov07b}. Therefore, the relative
amplitude of this dipole component must also be approximately constant
below $\sim 20$~keV and growing at higher energies, but the energy
dependence is expected to be different from that of the CG dipole. 

Unfortunately, within the RXTE energy band we cannot distinguish the CG
dipole from the large-scale structure one. Therefore, in our analysis
we estimated the CG dipole based on the known parameters of our 
motion with respect to the cosmic microwave background \citep{lineweaver96}. 

In Table \ref{dipoles} we present the measured CXB dipole in the energy 
band 3--8 keV and its value after subtraction of the CG component. 
We note that the obtained dipole amplitude and orientation are in
good agreement with the values determined from the analysis of 
HEAO1/A2 data \citep{scharf00}.

\begin{table}
\caption{Parameters of the CXB dipole component in the energy band 
3--8 keV determined from RXTE/PCA data} 
\begin{tabular}{lccc}
Map&Amplitude ($\Delta$, $10^{-3}$)& $l(^\circ)$ & $b(^\circ)$\\
\hline
All sky&$10.4\pm0.5$&$294\pm20$&$8\pm6$\\
CG corrected&$8.1\pm0.9$&$333\pm21$&$2\pm7$\\
\end{tabular}
\label{dipoles}
\end{table}

\subsection{X-ray emissivity of the local Universe}

In order to compare the CXB intensity variations with the expected
variations due to the nearby mass concentrations, we made use of the IRAS 
PSCz survey of galaxies \citep{saunders00}. 

The difference in the distribution of nearby ($<100$--200~Mpc) galaxies in 
a pair of directions will cause a difference in the CXB
intensities in these directions. Suppose that the X-ray emissivity per
unit number density of galaxies is constant ($\psi=dL/dN={\rm
  const}$), while the space density of galaxies  
($\rho=dN/dV$) varies with direction and with distance from us ($R$). 
Let the density of galaxies in one direction be $\rho_1(R)$
and in another $\rho_2(R)$. Then the difference of the CXB fluxes
$\Delta F$ created by the density variations, measured within a solid
angle $\Omega$ in these directions will be  

$$
\Delta F = \int_{0}^{R_{\rm max}}{{(\rho_1-\rho_2) \psi \Omega R^2 dR \over{4\pi R^2}}}={\Omega \psi \over{4\pi}} \int_{0}^{R_{\rm max}}{(\rho_1-\rho_2) dR}
$$

or

$$
\Delta F = {\Omega \psi \rho_0 \over{4\pi}} \int_{0}^{R_{\rm max}}{\left({\rho_1\over{\rho_0}}-{\rho_2\over{\rho_0}}\right) dR},
$$ 
where $\rho_0$ is the space density of galaxies averaged over a large volume.
The value $\psi \rho_0=\phi$ is the X-ray volume emissivity of the 
local Universe in units of \lum\ Mpc$^{-3}$. Substituting in the above
formula the average density $\rho_0$ for $\rho_2(R)$, we can construct a map of
predicted relative deviations of the CXB intensity from its average
level, generated by local ($\la150$ Mpc) inhomogeneities. 

We estimated the space density of IRAS galaxies with luminosities
$>10^{8} L_\odot$ from the PSCz catalog
by taking into account the dependence of the minimum detectable galaxy
luminosity on distance and the finite sky coverage of the survey.
We only considered density concentrations out to a radius of $\sim150$
Mpc.

\begin{figure}
\includegraphics[width=\columnwidth]{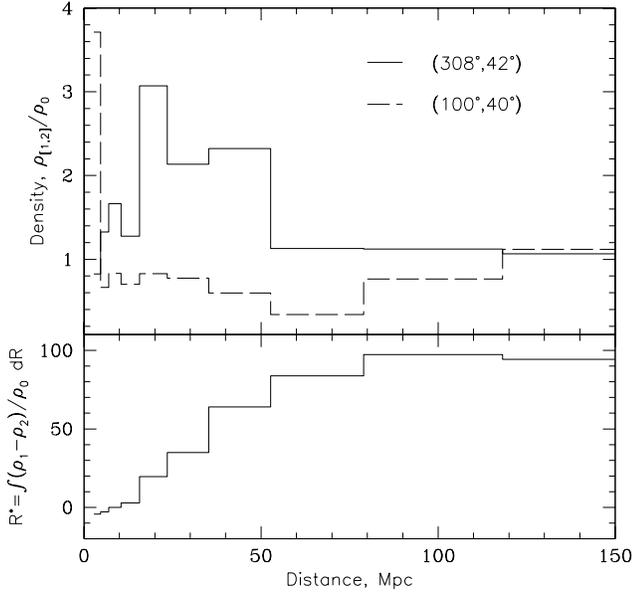}
\caption{Space density of IRAS galaxies calculated in two directions of the
sky within cones with a half-opening angle of 25$^\circ$ (upper panel) and the
radially integrated difference of these densities (lower panel).} 
\label{two_directions_iras}
\end{figure}

\begin{figure}
\includegraphics[width=\columnwidth]{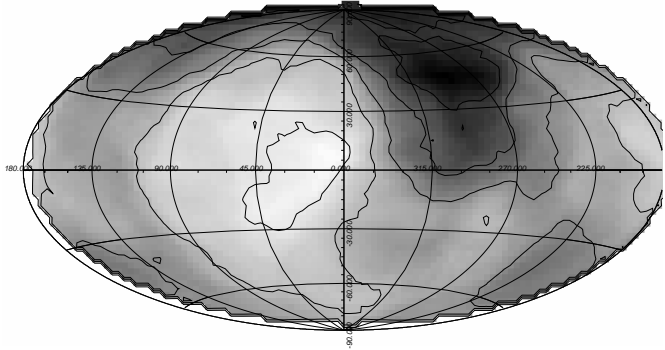}
\caption{Radially integrated relative space density of galaxies
($R^*=\int{(\rho_1-\rho_0})/\rho_0dR$), measured in cones with a
half-opening angle of $25^\circ$.}
\label{irmeasure}
\end{figure}

Fig.~\ref{two_directions_iras} shows the radially integrated difference
$\int(\rho_1-\rho_2)/\rho_0 dR$ of the IRAS galaxy number
densities in two particular directions, determined within cones with
a half-opening angle of 25$^\circ$. Fig.~\ref{irmeasure} shows an
all-sky map of relative galaxy number density variations  
$R^*=\int(\rho-\rho_0)/\rho_0 dR$, calculated in similar $25^\circ$
cones. It can be seen that mass 
concentrations in the nearby Universe create a clear excess in the
upper-right part of the map, which resembles the 
pecularity on the CXB intensity map (Fig.~\ref{map3_8}). Due to the
much smaller contrast of $R^*$ values in the southern Galactic
hemisphere, its correlation with the CXB intensity map provides very
little information in addition to the northern
hemisphere. Therefore, in studing the infrared-X-ray correlation we only 
considered the northern hemisphere.  

\begin{figure}
\includegraphics[width=\columnwidth,bb=0 90 594 689,angle=-90]{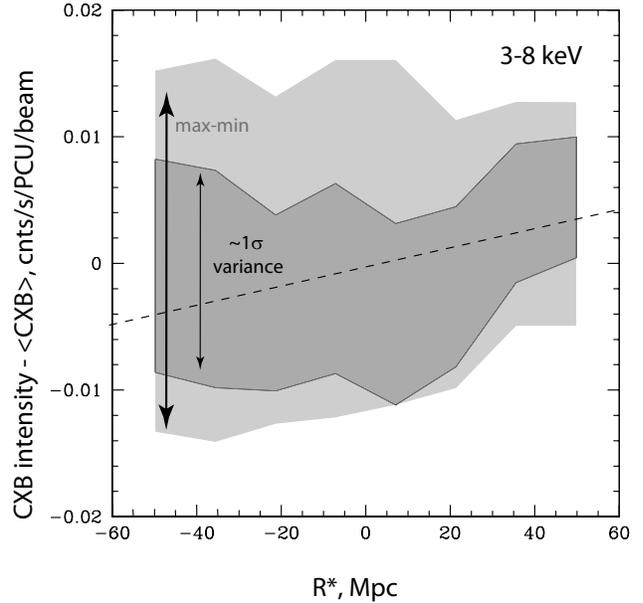}
\caption{Distribution of CXB variations measured by RXTE/PCA in the
energy band 3--8 keV in beams of size 1932~sq.~deg (cones with 
a $25^\circ$ half-open angle),  as a function of
radially integrated relative number densities of IRAS galaxies
($R^*=\int{(\rho-\rho_0 })/\rho_0dR$) calculated in the same
beams. The dashed line shows the best-fit linear correlation between  
these two quantities: $I$/(cnts/s/PCU/beam)$={\rm const}+
(7.7\pm7.0)\times10^{-5} R^*$/Mpc. }  
\label{correlation}
\end{figure}

In Fig.~\ref{correlation} we show the scatter plot of relative
variations of the CXB intensity (3--8 keV) with respect
to relative number density variations of nearby galaxies across the
sky. There is marginal evidence of a correlation between these
two quantities. Since we have filtered out sources with X-ray flux larger 
than $F_{\rm lim}$ from the all-sky map, the linear part of this
correlation provides us an estimate of the cumulative emissivity of 
sources with luminosities $L<L_{\rm lim}=4\pi F_{\rm lim} \langle
D\rangle ^2$, where $F_{\rm lim}$ is the typical sensitivity of the
survey, and $\langle D\rangle$ is the characteristic distance at which
inhomogeneities in the galaxy distribution are probed (see
e.g. Fig.~\ref{irmeasure}). We note that after we added new RXTE 
data, the typical sensitivity $F_{\rm lim}$ of our survey 
has slightly improved with respect to the value reported by
\cite{revnivtsev04} and is now $\sim 10^{-11}$  
\flux\ for Crab-like spectra in the energy band 2--10 keV. This translates
into a limiting luminosity $L_{\rm lim}\sim$1--3$\times 10^{42}$ \lum\ for 
$\langle D\rangle=30$--50 Mpc, the distances at which most of the local
anisotropy is created (see Fig.~\ref{two_directions_iras}). We thus
find from the linear part of 
the correlation shown in Fig.~\ref{correlation} that the cumulative
emissivity of local sources with $L<L_{\rm lim}$ is
$(4.4\pm3.9)\times10^{38}$ \lum\ Mpc$^{-3}$ in the energy band 2--10 keV.

On the other hand, the cumulative X-ray emissivity of nearby AGNs with
higher luminosities ($L\ga 10^{42}$~erg~s$^{-1}$) has been measured fairly
accurately by directly counting sources
\citep[e.g.][]{piccinotti82,ueda03,sazonov04}. Taking into account  
that in the 3--20~keV band the cumulative emission of such AGNs is
characterized by a spectral slope similar to 
that of the CXB, i.e. $\Gamma\sim 1.4$ \citep{sazonov07b}, 
we can convert the emissivity reported by
\cite{sazonov04} in the 3--20 keV band to 2--10 keV\footnote{Note that
we here take into account a recently realized correction for flux 
``leakage'' due to the fast motion of the PCA field of view during
RXTE slews: the fluxes of point sources reported by \cite{revnivtsev04}
were underestimated by a factor of 1/0.7.}: $dL/dV(\log
L>42)=(4.5\pm 0.9)\times10^{38}$ \lum\ Mpc$^{-3}$. 

Adding up the contributions of sources with luminosities below $\sim
10^{42}$~erg~s$^{-1}$ and AGNs with luminosities over $\sim
10^{42}$~erg~s$^{-1}$, estimated above, we find that the total X-ray 
emissivity of the local Universe in the energy band 2--10 keV is
$(9\pm4)\times10^{38}$ \lum\ Mpc$^{-3}$.

\subsection{Spectrum of variations}

The simple estimates presented in Sect.~\ref{subsec:systematics} suggest
that, due to the descrete nature of the CXB, variations of the CXB flux on
different angular scales should have different energy spectra. Indeed, 
on small angular scales (e.g. 1 sq.~deg), the Poisson variations of the
number of unresolved sources (with flux below the detection threshold)
dominate and one should expect that a few sources within the
field of view of the detector will provide a dominant contribution to
such variations. Therefore, the count rate variations will have
spectral hardness typical of such ($f\sim10^{-12-12.5}$ \flux) sources
and will have the amplitude give by eq.(\ref{poiss}). 

After filtering our bright sources and averaging the resulting count rates 
over areas of the sky $\gg 1$~sq.~deg, one should expect that the
spectrum of the variations will have a shape typical for the total
emissivity of the local Universe (Sazonov et al. 2007), since the
majority of the CXB flux variations on such angular scales are
produced by large-scale mass structures in the nearby Universe (note that
we have subtracted the estimated contributiuon of the CG dipole). 

\begin{figure}
\includegraphics[width=\columnwidth]{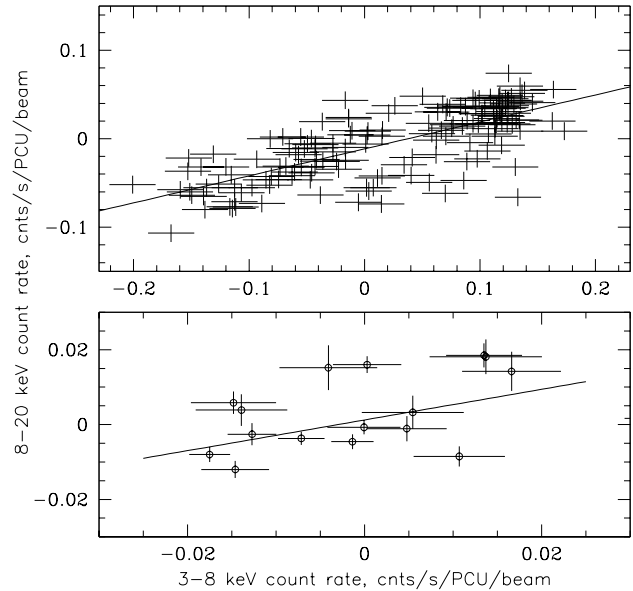}
\caption{Scatter plot of count rates in the energy bands 3--8 and
8--20~keV measured within the field of view of RXTE/PCA during its
observations of the 6 background pointings at different times 
(upper plot), and of fluxes averaged over different wide ($25^\circ$
half-opening angle) cones after application of all the data
filterings. The solid lines show the corresponding best-fit linear
correlations. Average CXB flux is subtracted from count rates in both
energy channels.}
\label{scatter_two_chans}
\end{figure}

As we show in Fig.~\ref{scatter_two_chans}, the slope of the scatter
plot of flux variations with respect to the average CXB intensity 
measured in PCA (1 sq.~deg)
beams in the 3--8 and 8--20 keV energy bands is approximately
$0.31\pm0.02$, which corresponds to the hardness ratio of a power-law
spectrum with a photon index $\Gamma=1.8\pm0.1$. For comparison, the
slope of the scatter plot of flux variations measured
over larger solid angles ($\sim2000$~sq.~deg) is $0.42\pm0.07$, which
corresponds to the 
hardness ratio of a power law with a photon index $\Gamma=1.4\pm0.2$. 
Therefore, we do find an indication that unresolved sources with flux
below the RXTE detection threshold have softer spectra than the CXB
(see a similar result obtained using GINGA data in \citealt{butcher97}).

Another important conclusion that can be drawn from this RXTE/PCA measurement
of the spectrum of large-angular-scale variations is that {\em the total
X-ray emission of the local Universe is characterized by spectral hardness
similar to that of the CXB}. This in particular implies that
emission from non-active galaxies and clusters of galaxies, which is
typically much softer than that from AGNs and especially from absorbed AGNs,
does not significantly contribute to the total X-ray emissivity of the local
Universe. Assuming that the cumulative emission spectrum of non-active
galaxies and low-luminosity ($L<10^{42}$~erg~s$^{-1}$ clusters of
galaxies is characterized by a hardness 
ratio (8--20 keV over 3--8 keV) that coresponds to a power law with a
photon index $\Gamma>2.5$ in the energy band 3--20 keV, we can
estimate that such objects contribute less than 15\% of the total
emissivity of the local Universe. This is consistent with our
estimates presented below that are based on the statistics of
local non-active galaxies and clusters of galaxies.

\subsection{Low-luminosity objects in the local Universe}

As we summarize in Table~\ref{budget} and detail below, it seems
possible to explain the total emissivity of sources with luminosities
${\rm \log L_{\rm 2-10~keV}}<42$, estimated above from the correlation
of the CXB anisotropies with the local large-scale structure, as a
superposition of contributions from known types of sources:
low-luminosity AGNs, clusters of galaxies, normal and starforming galaxies. 

\subsubsection{Low-luminosity AGN}

AGNs with X-ray luminosities below $10^{42}$~erg~s$^{-1}$, which we
refer to below as low-luminosity AGNs (LLAGNs) are known to be abundant 
in the local Universe \cite[e.g.][]{ho97}. \cite{elvis84} used a
sample of optically selected LLAGNs to estimate the X-ray luminosity
function of such objects by converting their measured $H_\alpha$
luminosities to X-ray luminosities under the assumption of linear correlation
between these two quantities. These authors concluded that the
cumulative emissivity of LLAGNs is $\sim 3\times10^{38}$ \lum Mpc$^{-3}$, 
i.e. comparable 
to that of AGNs with higher luminosities. 

\subsubsection{Clusters of galaxies}

The X-ray emission from clusters of galaxies is typically fairly soft,
especially from low-luminosity (low-mass) clusters. Therefore, the cumulative
emissivity of clusters of galaxies must strongly depend on the energy
band of the study. We consider here the standard X-ray energy band
2--10 keV, where the cumulative emissivity of clusters of galaxies can
be estimated from their soft X-ray luminosity function
\cite[e.g.][]{bohringer02,mullis04} using the well-known intracluster
gas temperature-luminosity relation (e.g. \citealt{markevitch98}). We
find that clusters with luminosities below $10^{42}$ \lum\
in the energy band 2--10 keV contribute together $\la10^{37}$
\lum\ Mpc$^{-3}$ to the local X-ray emissivity.

\subsubsection{Normal and starforming galaxies}

The contribution of normal and starforming galaxies to the X-ray
emissivity in the energy band 2--10 keV mainly comes from low-mass and
high-mass X-ray binaries (LMXBs and HXMBs), respectively. The shape of
the luminosity function of LMXBs is approximately constant while its
normalization is proportional to the galaxy mass
\cite[e.g.][]{fabbiano86,gilfanov04a}. Therefore, the cumulative emissivity of
normal galaxies can be estimated from the known value of 
integrated stellar mass density ($\Omega_*h\sim (1.6-3.5)\times10^{3}$,  
i.e. $\rho_*=(2.4-5.2)\times10^{8}\, M_\odot$ Mpc$^{-3}$) in the
low-redshift Universe \citep{cole01,kochanek01}. This gives a value of
(2--4)$\times10^{37}$ \lum\ Mpc$^{-3}$.

The emissivity of starforming galaxies in energy band 2-10 keV 
(via emission of HMXBs) can be
estimated either via the luminosity function  of nearby starforming
galaxies \cite[e.g.][]{ranalli05}, or by converting the average star
formation rate in the local Universe \cite[e.g.][]{gallego95} to an
X-ray volume emissivity using known statistical properties of HMXBs in
starforming galaxies  \cite[e.g.][]{gilfanov04}. Both types of
estimate suggest that the emissivity is approximately
(3--4)$\times10^{37}$  \lum\ Mpc$^{-3}$. 

\begin{table}
\caption{Emissivities (2--10~keV) of different types of low-luminosity objects
 ($L_{\rm 2-10~keV}<10^{42}$~\lum) in the local Universe in units of 
$10^{38}$ \lum\ Mpc$^{-3}$}
\tabcolsep=0.7cm
\begin{tabular}{lc}
Class&$dL/dV$\\
\hline
LLAGN&$\sim3$\\
Clusters of galaxies&$\la0.1$\\
Starforming gal.(HMXBs)&0.2--0.3\\
Normal gal.(LMXBs)&0.2--0.4\\
\hline
Sum of the above&$\sim4$\\
\hline
This paper, $(\log L_{\rm 2-10~keV}\la42)$&$4.4\pm3.9$\\
\end{tabular}
\label{budget}
\end{table}

\section{Summary}

Using data collected by the PCA instrument during slews of the RXTE
observatory between its pointed observations in 1996--2007 we have
obtained an all-sky map of the cosmic X-ray background
intensity. Through a careful data analysis we measured the CXB
intensity variations with an accuracy $\sim 0.5$\%.

On small angular scales we detected variations with $\sim7$\% FWHM, which 
are most likely caused by the Poisson variations of the number of weak sources 
below the detection threshold in the field of view of the detector.

Upon application of different filters, we detected statistically
significant variations of the CXB intensity on the 20--40$^\circ$ angular
scales with an amplitude $\sim2$\%. Part of these variations is
correlated with mass concentrations in the nearby ($D<150$ Mpc)
Universe, which allowed us to make an estimate of the total emissivity of
low-luminosity ($\log L_{\rm 2-10~keV}\la42$) sources. The obtained value is in
agreement with those obtained previously based on HEAO1/A2 data
\citep{jahoda91,miyaji94}. 

The spectral hardness ratio of large-angular-scale variations of the
CXB intensity is compatible with that of the average CXB
spectrum. This allowed us to put an upper limit on the combined
contribution of soft X-ray sources (normal and starforming galaxies,
and low-mass clusters of galaxies) to the total emissivity of
the local Universe in the energy band 2--10 keV: $<15$\%. Most of the
observed CXB anisotropy can be attributed to low-luminosity AGNs.

\begin{acknowledgements}
Authors thank Rashid Sunyaev for valuable discussions.
MR thank Gerard Lemson for assistance in the usage of the results of the 
Millennium simulations. This research made use of data obtained from the High 
Energy Astrophysics Science Archive Research Center Online Service,
provided by the NASA/Goddard Space Flight Center. This work was
supported by DFG-Schwerpunktprogramme SPP 1177.

\end{acknowledgements}

\end{document}